\documentclass[amssymb,amsmath,aps,prl,twocolumn,superscriptaddress,showpacs]{revtex4-1}
\usepackage{graphicx}
\usepackage{dcolumn}
\usepackage{bm}
\usepackage{multirow}
\usepackage{color}

\begin{document}

\title{Thermal Control of Spin Excitations in the Coupled Ising-Chain
Material RbCoCl$_3$}

\author{M. Mena}
\affiliation{London Centre for Nanotechnology and Department of Physics and 
Astronomy, University College London, Gower Street, London WC1E 6BT, United 
Kingdom}
\affiliation{Laboratory for Neutron Scattering and Imaging, Paul Scherrer 
Institute, CH-5232 Villigen PSI, Switzerland}
\author{N. H\"anni}
\affiliation{Department of Chemistry and Biochemistry, University of Bern, 
Freiestrasse 3, CH-3012 Bern, Switzerland}
\affiliation{Physikalisches Institut, University of Bern, Siedlerstrasse 5, 
CH-3012 Bern, Switzerland}
\author{S. Ward}
\affiliation{Laboratory for Neutron Scattering and Imaging, Paul Scherrer 
Institute, CH-5232 Villigen PSI, Switzerland}
\affiliation{Department of Quantum Matter Physics, University of Geneva, 
CH-1211 Geneva 4, Switzerland}
\author{E. Hirtenlechner}
\affiliation{Laboratory for Neutron Scattering and Imaging, Paul Scherrer 
Institute, CH-5232 Villigen PSI, Switzerland}
\affiliation{Institut Laue Langevin, CS 20156, F-38042 Grenoble, France}
\author{R. Bewley}
\affiliation{ISIS Facility, STFC Rutherford Appleton Laboratory, Harwell 
Campus, Didcot OX11 0QX, United Kingdom}
\author{C. Hubig}
\affiliation{Arnold Sommerfeld Center for Theoretical Physics, 
Ludwig-Maximilians-University Munich, 80333 M{\"u}nchen, Germany}
\affiliation{Max-Planck-Institut f\"ur Quantenoptik, 85748 Garching, Germany}
\author{U. Schollw{\"o}ck}
\affiliation{Arnold Sommerfeld Center for Theoretical Physics, 
Ludwig-Maximilians-University Munich, 80333 M{\"u}nchen, Germany}
\author{B. Normand}
\affiliation{Paul Scherrer Institute, 
CH-5232 Villigen PSI, Switzerland}
\author{K. W. Kr{\"a}mer}
\affiliation{Department of Chemistry and Biochemistry, University of Bern, 
Freiestrasse 3, CH-3012 Bern, Switzerland}
\author{D. F. McMorrow}
\affiliation{London Centre for Nanotechnology and Department of Physics and 
Astronomy, University College London, Gower Street, London WC1E 6BT, United 
Kingdom}
\author{Ch. R\"uegg}
\affiliation{Department of Quantum Matter Physics, University of Geneva, 
CH-1211 Geneva 4, Switzerland}
\affiliation{Paul Scherrer Institute, 
CH-5232 Villigen PSI, Switzerland}
\affiliation{Institute for Quantum Electronics, ETH Z\"urich, CH-8093 Z\"urich, Switzerland}
\affiliation{Institute of Physics, Ecole Polytechnique Federale de Lausanne, CH-1015 Lausanne, Switzerland}

\date{\today}

\begin{abstract}
We have used neutron spectroscopy to investigate the spin dynamics of 
the quantum ($S = 1/2$) antiferromagnetic Ising chains in RbCoCl$_3$. The 
structure and magnetic interactions in this material conspire to produce two 
magnetic phase transitions at low temperatures, presenting an ideal opportunity 
for thermal control of the chain environment. The high-resolution spectra we
measure of two-domain-wall excitations therefore characterize precisely both 
the continuum response of isolated chains and the ``Zeeman-ladder'' bound 
states of chains in three different effective staggered fields in one and 
the same material. We apply an extended Matsubara formalism to obtain a 
quantitative description of the entire dataset, Monte Carlo simulations to 
interpret the magnetic order, and finite-temperature DMRG calculations to 
fit the spectral features of all three phases.
\end{abstract}

\maketitle

Quantum systems that display one-dimensional (1D) nature \cite{giamarchi2004}
and Ising exchange \cite{ising1925} exhibit a very rich variety of phenomena. 
These include a gapped excitation continuum \cite{Yoshizawa1981,Nagler1983,
Nagler1983_2}, excited bound states with emergent E$_8$ symmetry 
\cite{coldea2010}, quantum criticality \cite{sachdev1999,ronnow2005,
kraemer2012}, and topological excitations \cite{faure2017}. Recent intense 
interest in coupled Ising-chain physics was sparked by the ferromagnetic (FM) 
material CoNb$_2$O$_6$ \cite{coldea2010,morris2014,robinson2014,cabrera2014}, 
whose spectrum of magnetic excitations is divided between a kinetic bound 
state and a two-domain-wall continuum, which itself splits into confined 
(E$_8$) bound states. The antiferromagnetic (AF) ACo$_2$V$_2$O$_8$ (A \! = \! 
Ba, \! Sr) compounds \cite{kimura07,kimura13,bera14,wang2015} have spiralling 
Ising chains with significant Heisenberg interactions, not only separating the
transverse and longitudinal bound-state excitations \cite{grenier2015,bera17}
but also realizing both uniform and staggered, longitudinal and transverse 
applied fields \cite{kimura13} in which to measure different types of bound 
state and critical behavior \cite{faure2017,wang16,matsuda17,wang18}. However, 
both the response of a truly decoupled Ising chain and distinguishing the 
temperature dependences of all these features remain as challenging problems.

\begin{figure}[t]
\includegraphics[width=8.78cm]{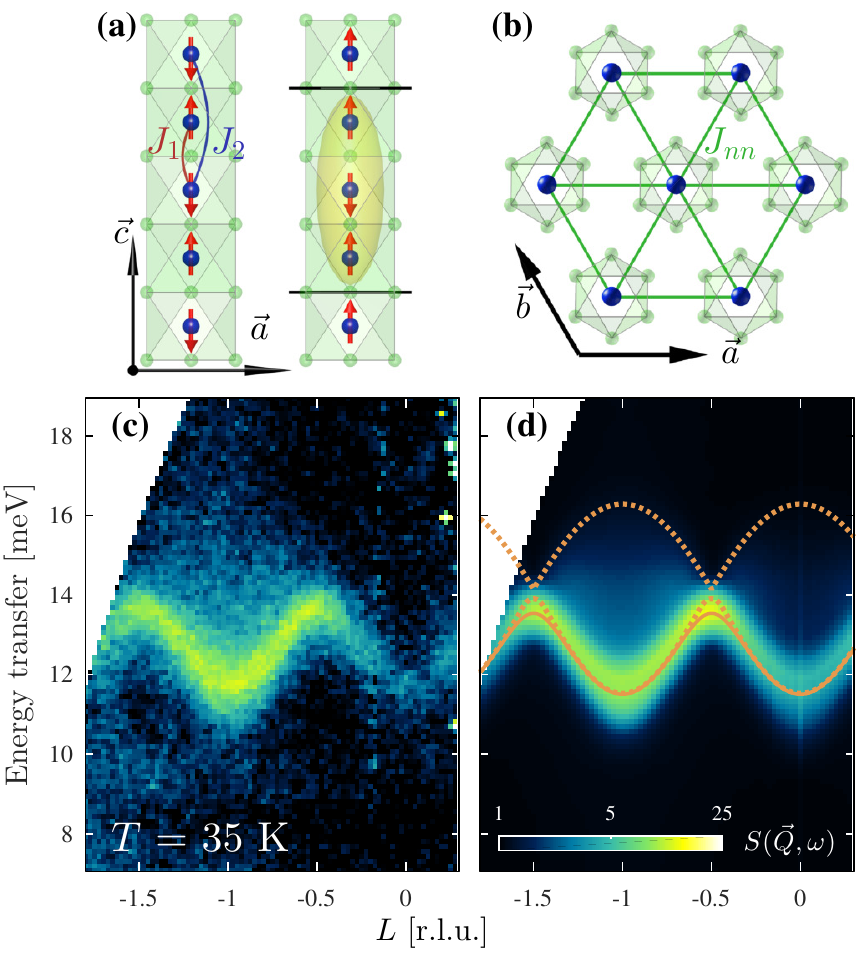}
\caption{\label{fig1}(a) Schematic representation of chains in RbCoCl$_3$. 
The Co$^{2+}$ ions (blue) are embedded in Cl$^{-}$ antiprisms (light green), 
an environment in which their ground state is an Ising doublet. The resulting 
effective $S = 1/2$ spins (red arrows) order antiferromagnetically along the 
$c$ axis. The yellow ellipse denotes a pair of interacting domain walls. 
(b) Triangular lattice of chains in RbCoCl$_3$. (c) Scattered intensity, 
$S(\vec{Q},\omega)$, measured at 35 K. (d) $S(\vec{Q},\omega)$ calculated by 
combining DMRG results with extended Matsubara analysis (see text). Dotted 
orange lines mark the edges of the continuum; the solid orange line marks 
the bound mode that separates from the continuum due to the interaction 
$J_2$ [panel (a)].} 
\end{figure}

A paradigm for Ising-chain physics is provided by the ACoX$_3$ hexagonal 
perovskites [space group P$6_3$/mmc, shown in Figs.~\ref{fig1}(a) and 
\ref{fig1}(b)], which include CsCoCl$_3$ \cite{Lehmann1981,Yoshizawa1981,
Nagler1983,Matsubara1991,goff1995}, CsCoBr$_3$ \cite{Lehmann1981,Nagler1983,
Nagler1983_2,Matsubara1991}, RbCoCl$_3$ \cite{Lockwood1983,Matsubara1991}, 
and TlCoCl$_3$ \cite{oosawa2006}. Their in-chain interactions are AF and 
their spectra provided early examples of the continuum arising from pairs 
of moving domain walls (also referred to as ``kinks'' and ``solitons''). 
Their special feature is that the Ising chains form a triangular lattice 
[Fig.~\ref{fig1}(b)], frustrating the AF interchain interactions. They 
typically show two magnetic ordering transitions, which in RbCoCl$_3$ occur 
at $T_{N1} = 28$ K and $T_{N2} = 12$ K \cite{haenni2017}. Thus, in contrast 
to CoNb$_2$O$_6$ and ACo$_2$V$_2$O$_8$, different chains experience different 
environments at different temperatures. Treating the neighboring chains as an 
effective staggered field, when this is zero one expects the spectrum of the 
isolated chain, a continuum with cosinusoidal boundaries \cite{ishimura1980}, 
whereas for finite staggered fields the spectrum is a Zeeman ladder of bound 
states \cite{shiba1980}. 

In this Letter, we study the magnetic excitations of RbCoCl$_3$ in 
unprecedented detail by combining state-of-the-art instrumentation for 
neutron spectroscopy with a systematic thermal control of the different 
effective-field contributions. This allows an unambiguous separation of 
isolated-chain and staggered-field physics. Rising temperature causes both 
a band-narrowing and a broadening of bound and continuum features, as well 
as strong changes to the interchain coupling effects. These we explain by 
combining an extended Matsubara description with finite-temperature, 
time-dependent density-matrix renormalization-group (DMRG) calculations.

Three high-quality RbCoCl$_3$ single crystals of total mass 7.6 g were grown 
by the Bridgman technique in a moving vertical furnace \cite{haenni2017} and 
coaligned in the (\textit{HHL}) plane. Inelastic neutron scattering 
experiments were performed on the direct-geometry time-of-flight spectrometer 
LET (ISIS, UK) \cite{bewley2011}. High-statistics data were collected with 
incident energy $E_i = 25$ meV and (00$L$) perpendicular to the incident beam, 
at temperatures of 4, 18, and 35 K, i.e.~in each magnetically ordered phase 
and above both. Lower-statistics data were collected at 8 K, 10.5 K, and 23 K 
with $E_i$ = 20 meV. The data were corrected for detector efficiency and 
outgoing-to-incoming wave-vector ratio, $k_f / k_i$, using the program MANTID 
\cite{mantid}. Data sets for the scattered intensity, $S(\vec{Q},\omega)$, 
were analyzed with the HORACE software package \cite{ewings2016}. 

\begin{figure}[t]
\includegraphics[width=8.78cm]{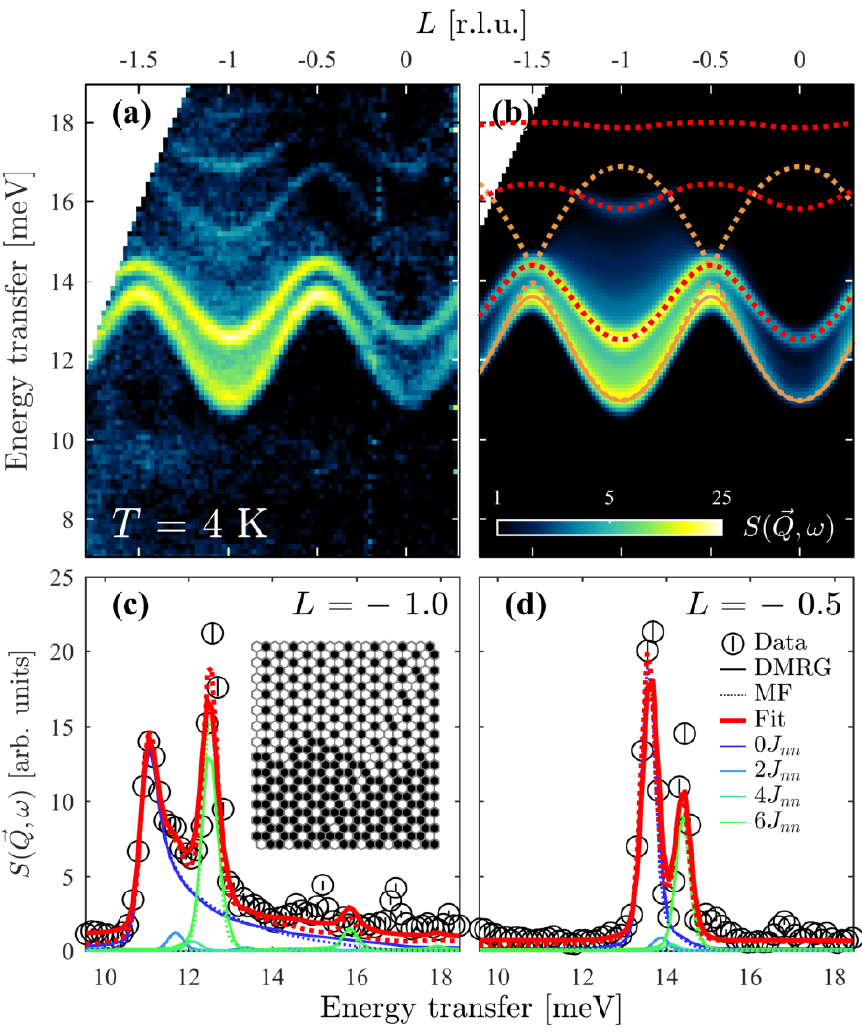}
\caption{\label{fig4K} (a) $S(\vec{Q},\omega)$ measured at 4 K. (b) $S(\vec{Q},
\omega)$ calculated in Matsubara formalism (MF) with the parameters in the 
text. Orange lines as in Fig.~1(d). Red lines show the bound-state excitations 
of the Zeeman ladder. (c) and (d) $S(\vec{Q},\omega)$ at 4 K (black symbols) 
integrated over windows $-1.05 < L < -0.95$ (c) and $-0.55 < L < -0.45$ (d). 
Red lines show optimized fits obtained by summing contributions from chains 
in different effective fields (blue-to-green lines). These panels are enlarged 
in Fig.~S5 of the SM \cite{sm}. The inset in panel (c) shows a typical planar 
spin configuration, dominated by FI domains, obtained from CHB Monte Carlo 
simulations at $T < T_{N2}$; black hexagons denote up-spins and white 
down-spins.}
\end{figure}

\begin{figure}[t]
\includegraphics[width=8.78cm]{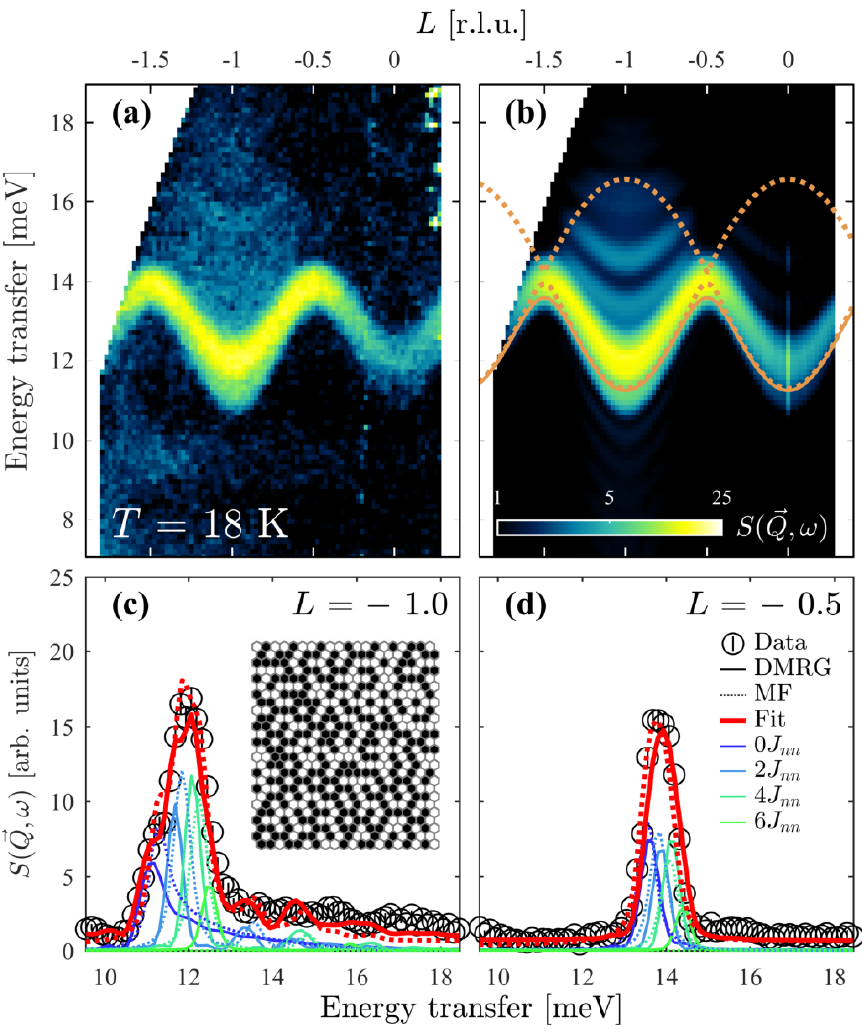}
\caption{\label{fig18K} (a) $S(\vec{Q},\omega)$ measured at 18 K. (b) 
$S(\vec{Q},\omega)$ calculated by DMRG. Orange lines as in Fig.~1(d). (c) and 
(d) Integrated scattered intensities following the conventions of Figs.~2(c) 
and 2(d). These panels are enlarged in Fig.~S5 \cite{sm}. The inset in panel 
(c) shows a typical planar spin configuration in the PDAF phase from CHB 
Monte Carlo.} 
\end{figure}

\begin{figure}[t]
\includegraphics[width=8.78cm]{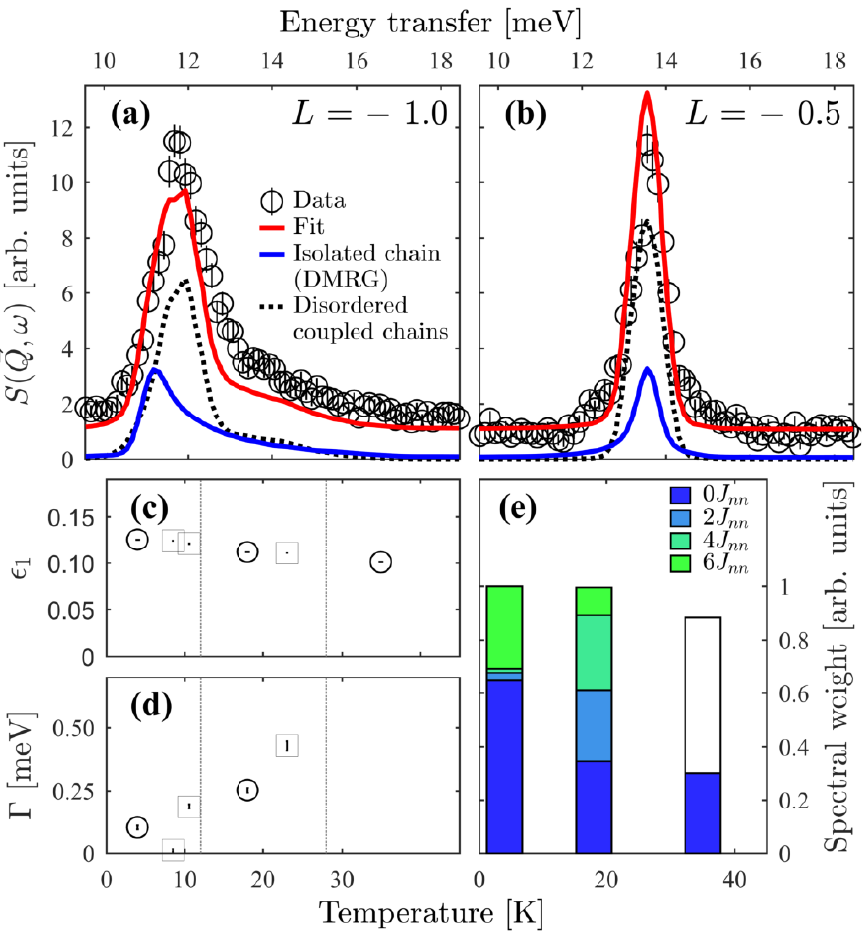}
\caption{\label{fig35K} (a) and (b) Integrated scattered intensities at 35 K 
following the conventions of Figs.~2(c) and 2(d). Red lines show fits 
combining DMRG for the isolated chain (solid blue) and Matsubara-framework 
modelling of coupled chains with static thermal disorder (dashed black lines). 
(c) Effective $\epsilon_1(T)$, with circles (squares) from fits to high- 
(low-)statistics datasets. (d) Effective Lorentzian width, $\Gamma (T)$. 
(e) Integrated intensity associated with each staggered field at 4 K and 
18 K, taken from DMRG fits; at 35 K we distinguish the intensity (blue) 
associated with the isolated-chain response from that (white) due to chains 
in thermally disordered effective fields.}
\end{figure}

We begin with an overview of the experimental data, which are shown in 
Fig.~\ref{fig1}(c), Figs.~\ref{fig4K}(a,c,d), Figs.~\ref{fig18K}(a,c,d), 
and Figs.~\ref{fig35K}(a,b). First, we identify a band of 
excitations  in the one-spin-flip (two-domain-wall) sector with a spin gap 
of 11 meV and a cosinusoidal dispersion [Fig.~\ref{fig1}(c)]. Second, our 
measurements confirm the strongly 1D nature of RbCoCl$_3$, as shown in Sec.~S1 
of the Supplemental Material (SM) \cite{sm}. All data presented here were 
therefore averaged over $H$ and $K$ and binned as functions of the energy 
transfer and the momentum along $L$. In detail, the width of the band we 
observe at 35 K [Fig.~\ref{fig1}(c)] far exceeds the instrumental resolution 
and its line shape [Fig.~\ref{fig35K}(a)] suggests a continuum. The spectrum 
sharpens at 18 K [Fig.~\ref{fig18K}(a)], presenting more lines with more 
intensity and less broadening [Figs.~\ref{fig18K}(c) and \ref{fig18K}(d)]. 
At 4 K, the scattered intensity splits clearly into two different types of 
feature [Fig.~\ref{fig4K}(a)], a continuum [Fig.~\ref{fig4K}(c)] whose maximum 
disperses between 11 and 13.5 meV and a set of resolution-limited modes, 
the Zeeman ladder, whose lowest member [Fig.~\ref{fig4K}(d)] disperses 
between 12.5 and 14.5 meV. It is evident that thermal effects go beyond a 
simple line-broadening and include effective control of the staggered field 
due to the different 3D ordered phases. 

To illustrate the influence of magnetic order on the measured Ising-chain 
dynamics, we have implemented a Cluster Heat Bath (CHB) Monte Carlo algorithm, 
as detailed in Sec.~S2 of the SM \cite{sm}. Developed originally for studies 
of ACoX$_3$ compounds \cite{matsubara1997,koseki1997,koseki2000}, CHB 
simulations below $T_{N2}$ show well-ordered chains aligning to form domains 
of ferrimagnetic (FI) ``honeycomb'' planar order \cite{mekata1977,koseki2000}, 
represented in the inset of Fig.~\ref{fig4K}(c). Interchain interactions are 
dominated by the nearest-neighbor (nn) AF coupling, $J_{nn}$ 
[Fig.~\ref{fig1}(b)], whose frustration leads to a massive degeneracy, 
and the FI order is selected by a weak FM next-nearest-neighbor (nnn) 
term, $J_{nnn}$ \cite{mekata1977}. The partially disordered AF (PDAF) 
\cite{mekata1977,goff1995} state between $T_{N2}$ and $T_{N1}$ is characterized 
primarily by a decrease in FI interchain order [inset, Fig.~\ref{fig18K}(c)], 
while the staggered magnetization within each chain is also impacted weakly 
\cite{haenni2017} by thermally excited domain walls \cite{todoroki2003}. 
Although 3D order is lost above $T_{N1}$, both susceptibility and diffuse 
scattering measurements \cite{haenni2017} suggest that anomalously slow 
short-range correlations persist up to 60--80 K, and we will quantify this 
effect.

The Hamiltonian of a single Ising-Heisenberg chain with nn and nnn in-chain 
interactions [Fig.~\ref{fig1}(a)] is
\begin{eqnarray}
\mathcal{H} & = & \sum_j 2 J_1 [ S^z_j S^z_{j+1} + \epsilon_1 (S^x_j S^x_{j+1} + 
S^y_j S^y_{j+1})] + h_j S^z_j \nonumber \\ & & \;\;\;\;\;\;\;\; + 2 J_2 [ S^z_j 
S^z_{j+2} + \epsilon_2 (S^x_j S^x_{j+2} + S^y_j S^y_{j+2})], 
\label{eh}
\end{eqnarray}
where $|J_2| \ll J_1$, $\epsilon_1 \ll 1$, and $h_j$ is the effective staggered 
field due to magnetic order. For this class of compounds, $J_2$ is thought to 
be FM \cite{Matsubara1991}. An alternative approach \cite{goff1995} using only 
nn terms and including an anisotropic splitting of the Co$^{2+}$ Kramers 
doublet is also thought \cite{shiba2003} to provide a complete treatment of 
the orbital terms. 

In this formalism, which we term the ``Matsubara framework,'' the matrix 
elements of Eq.~(\ref{eh}) expressed in terms of the separation, $\nu$, of 
two domain walls are  
\begin{eqnarray}
\langle 1 | \mathcal{H}(\vec{Q}) | 1 \rangle & = & 2J_1 ( 1 + \epsilon_1^2)
 + 2 J_2 [ 1 - \epsilon_2 \cos{(2 \pi L)}] + h, \nonumber \\ 
\langle \nu | \mathcal{H}(\vec{Q}) | \nu \rangle & = & 2J_1 ( 1 + 
\tfrac{3}{2}\epsilon_1^2) + 4 J_2 + \nu h, \nonumber 
\end{eqnarray}
\begin{eqnarray}
\langle \nu | \mathcal{H}(\vec{Q}) |\nu \pm 2 \rangle & = & \epsilon_1 J_1 \, 
(1 + e^{\mp 2i \pi L}), \nonumber \\
\langle \nu | \mathcal{H}(\vec{Q}) | \nu \pm 4 \rangle & = & -\tfrac{1}{2} 
\epsilon_1^2 J_1 \, (1 + e^{\mp4 i \pi L}),
\end{eqnarray}
and $\langle \nu | \mathcal{H}(\vec{Q}) | \nu' \rangle = 0$ otherwise 
\cite{ishimura1980,Matsubara1991,shiba2003}. As noted above, the spectrum 
of this Hamiltonian forms a continuum when $h = 0$ [dashed orange lines in 
Figs.~\ref{fig1}(d), \ref{fig4K}(b), and \ref{fig18K}(b)] and a Zeeman ladder 
when $h > 0$. With $h = 0$ but $J_2$ non-zero, an additional bound mode 
separates from the continuum around half-integer values of $L$ [solid orange 
lines], while the intensity in the continuum shifts towards its lower edge. By 
calculating the spectral weights from the Green function \cite{ishimura1980}, 
the full dynamical structure factor shown in Fig.~\ref{fig4K}(b) was computed 
using Eq.~(\ref{eh}) with the parameters discussed below. 

Because $J_{nnn}$ is much smaller than the instrumental resolution, at 4 K 
we model the effective staggered field by taking $h = m J_{nn}$ with $m \in 
\{0,2,4,6\}$. For perfect FI order, 2/3 of the sites experience a vanishing 
field, $h = 0$, and 1/3 a field $h = 6J_{nn}$; in practice, planar domain 
boundaries between different FI regions [inset Fig.~\ref{fig4K}(c)] produce 
small numbers of chains subject to the two intermediate fields. At 18 K, 
the PDAF regime presents a mix of all four staggered fields [inset 
Fig.~\ref{fig18K}(c)]. At 35 K, more detailed considerations are 
required as the chain environment becomes increasingly random.

Concerning in-chain dynamics at finite temperatures, the domain-wall pair 
excited by the scattered neutron is itself scattered by thermally excited 
domain walls \cite{koseki1997,todoroki2003}. This increases the effective 
localization of both continuum and bound modes, as well as reducing their 
lifetimes, which changes the line shapes into Voigt functions with a 
constant Gaussian width of $0.32$ meV (FWHM instrument resolution) and a 
temperature-dependent Lorentzian width, $\Gamma(T)$, which we take to be 
$\vec{Q}$-independent. In the Matsubara framework we model the band-flattening 
due to the increased localization by a temperature-driven decrease of 
$\epsilon_1$ in Eq.~(\ref{eh}) \cite{James2009}, taking $\epsilon_2$ as a 
constant. Below we compute the spectral functions at all temperatures from 
DMRG calculations based only on the low-$T$ parameters, and use the effective 
$\Gamma(T)$ and $\epsilon_1(T)$ as an aid to physical interpretation. 

A fully quantitative account of continuum and bound-state energies, of 
line shapes, and of the overall scattered intensities [color contours in 
Figs.~\ref{fig4K}(b), \ref{fig18K}(b), and \ref{fig1}(b)] is obtained with 
the parameter values $J_1 = 5.89(1)$ meV, $J_2 = - 0.518(1)$ meV, $J_{nn} = 
0.129(1)$ meV, and $\epsilon_2 = 0.605(1)$, valid at all temperatures, and 
with $\epsilon_1(T) = 0.126(1)$ at 4K, 0.112(1) at 18 K, and 0.101(2) at 35 
K [Fig.~\ref{fig35K}(c)]; $\Gamma(T) = 0.102(8)$ meV at 4 K and 0.25(1) meV 
at 18 K [Fig.~\ref{fig35K}(d)], and is not defined at 35 K (below). Figure 
\ref{fig35K}(e) shows the optimized weights associated with chains in 
different effective fields at each temperature.  

The separation of isolated-chain from staggered-field physics is clearest 
at 4 K [Fig.~\ref{fig4K}(a)]. The former is responsible for the 11 meV peak 
in Fig.~\ref{fig4K}(c), whose continuum tail extends to 17 meV, and the 
latter, with $h = 6J_{nn}$, for the sharp mode at 12.5 meV. The separation of 
these two primary contributions is emphasized by the small $\Gamma$ and the 
near-absence of the intermediate staggered fields. We fit the relative weights 
of the staggered fields as 67(3), 3(3), 0(3), and 31(3)\% for $h$ from 0 to 
$6J_{nn}$, and thus the 0 and $6J_{nn}$ components are rather close to the 
ideal 2:1 ratio of the FI phase. 

The most marked difference at 18 K [Fig.~\ref{fig18K}] is the strong shift 
towards contributions from the staggered fields $2J_{nn}$ and $4J_{nn}$, whose 
spectra overlap due both to their proximity in energy and to the increased 
$\Gamma$. The populations of the four staggered fields are 39(5), 35(5), 27(5), 
and 0(5)\%, which lie close to our optimal CHB Monte Carlo results of 39, 35, 
20, and 6\%.

Finally, at 35 K one may expect interchain correlations to be lost and 
the scattering to be dominated by isolated-chain physics, as suggested 
by Fig.~\ref{fig1}(c). However, in Fig.~\ref{fig35K}(a) it is clear that 
the isolated chain cannot provide an acceptable fit. To model the effects of 
thermal disorder, both in localizing the in-chain dynamics and in randomizing 
the chain environment, we have conducted Monte Carlo simulations of different 
static domain-wall distributions within the Matsubara framework, as described 
in Sec.~S5 of the SM \cite{sm}. Qualitatively, we find a broadening dictated 
by $2J_{nn}$, that becomes independent of the chain-length distribution, and 
a shift of weight towards the band center that again is a localization effect 
reproduced using $\epsilon_1$. Quantitatively, however, a definitive fit of 
the 35 K data would require a systematic account of domain-wall dynamics.

For such a microscopic analysis of combined quantum and thermal fluctuations 
in the Ising chain, we have performed time-dependent DMRG calculations 
\cite{schollwoeck11, hubig17:_gener} to obtain the finite-temperature spectral 
functions, $S(\vec{Q},\omega,T)$, as described in Sec.~S3 of the SM \cite{sm}. 
The spatial and temporal evolution of a spin-flip excitation was computed for 
512-site chains and a linear prediction method \cite{barthel09:_spect} used 
to access long effective times despite rapid entanglement growth at the higher 
temperatures. At 4 K we obtain the optimized DMRG parameters $J_1 = 5.86$ meV, 
$J_2 = -0.576$ meV, $J_{nn} = 0.128$ meV, $\epsilon_1 = 0.126$, and $\epsilon_2
 = 0.559$. By including the different staggered fields and the instrumental 
resolution, we obtain the spectrum shown in Fig.~\ref{fig4K}(b), where the 
DMRG and extended Matsubara results are indistinguishable. For a quantitative 
visualization of the accuracy with which the experimental peaks and line 
shapes are reproduced, the DMRG results for all staggered fields are also 
shown as the solid blue-to-green lines in Figs.~\ref{fig4K}(c) and 
\ref{fig4K}(d), and in detail in Sec.~S4 \cite{sm}. 

DMRG captures all of the thermal broadening effects arising from domain-wall 
scattering in a single chain. As Fig.~\ref{fig18K} makes clear, this provides 
an excellent account of the 18 K spectrum in all its details, although the 
fitted chain population distribution of 35, 27, 28, and 10\% is somewhat 
different from the less-constrained Matsubara fit (Sec.~S4 \cite{sm}). At 
35 K, where the approximation of chains in a well-ordered staggered field is 
no longer appropriate, we use our DMRG results for the scattered intensity, 
$I_0 (\omega)$, of the isolated chain to reproduce the contributions of chains 
decoupled from their neighbors by thermal fluctuations. In combination with 
the contribution of chains coupled by the random fields of a thermal 
distribution of static domain walls, which as above we model in the Matsubara 
framework, our optimal fit to the measured intensity [Figs.~\ref{fig35K}(a) 
and \ref{fig35K}(b)] reveals that approximately 63(5)\% of the spectral weight, 
marked in white in Fig.~\ref{fig35K}(e), is contributed by chain segments in 
non-vanishing effective fields (Sec.~S5 \cite{sm}). Thus we quantify the 
extent to which interchain correlations remain important at $T > T_{N1}$ 
\cite{haenni2017}, i.e.~close to but above the regime of finite long-range 
order.

We comment that higher modes of the Zeeman ladder are observed in the spectrum 
between 15 and 18 meV. These are rather sharp at 4 K [Figs.~\ref{fig4K}(a) and 
\ref{fig4K}(c)] but broad and weak at 18 K [Fig.~\ref{fig18K}(a)]. In our 
Matsubara and DMRG results [Fig.~\ref{fig4K}(b)], these modes are present with 
approximately the measured position, but their dispersion and intensity cannot 
be fitted with the same quantitative accuracy as the other spectral features.
This suggests that higher-energy corrections \cite{goff1995} or an RPA-type 
extension of the Matsubara formalism may be required. 

To summarize, we have studied both the continuum and the bound-state 
excitations of the quasi-1D AF Ising chain in a single material, RbCoCl$_3$. 
By using temperature to control the type of 3D magnetic order, we measure 
the dynamical response in different effective magnetic fields. We model the 
continuum of the isolated chain, the Zeeman ladders in all staggered fields, 
and the broad response in thermally randomized fields by an analytical 
domain-wall formalism and by DMRG, to obtain a quantitative description 
of the spectrum in each of the temperature regimes investigated. Our results 
constitute a frontier in exploring the finite-temperature response of quantum 
spin systems and highlight the need for systematic theoretical methods to 
treat this problem in 3D. 

We are grateful to R. Coldea and F. Essler for helpful discussions. We thank 
the neutron scattering facilities ISIS (Rutherford Appleton Laboratory, UK), 
SINQ (Paul Scherrer Institute, Switzerland), and the Institut Laue-Langevin 
(France). This research was supported by the UK Engineering and Physical 
Sciences Research Council (EPSRC) under Grant EP/N027671/1, by the European 
Research Council (ERC) under the EU Horizon 2020 research and innovation 
program Grants No.~681654 (HyperQC) and No.~742102 (QUENOCOBA), by the 
Bavarian Elite Network ExQM, and by the Swiss National Science Foundation 
(SNF) under Grants No.~200020-132877 and No.~200020-150257.

\clearpage

\setcounter{figure}{0}
\renewcommand{\thefigure}{S\arabic{figure}}

\setcounter{equation}{0}
\renewcommand{\theequation}{S\arabic{equation}}

\setcounter{section}{0}
\renewcommand{\thesection}{S\arabic{section}}

\setcounter{table}{0}
\renewcommand{\thetable}{S\arabic{table}}

\onecolumngrid

\vskip1cm

\centerline{\large {\bf {Supplemental Material for "Thermal Control of Spin Excitations }}} 

\vskip1mm

\centerline{\large {\bf {in the Coupled Ising-Chain Material RbCoCl$_3$"}}}

\vskip4mm

\centerline{M. Mena, N. H\"anni, S. Ward, E. Hirtenlechner, R. Bewley, 
C. Hubig,} 

\centerline{U. Schollw{\"o}ck, B. Normand, K. W. Kr{\"a}mer, D. F. McMorrow, 
and Ch.~R\"uegg}

\vskip8mm

\twocolumngrid

\subsection{S1. One-Dimensional Nature of RbCoCl$_3$}

\begin{figure}[b]
\includegraphics[width=8.78cm]{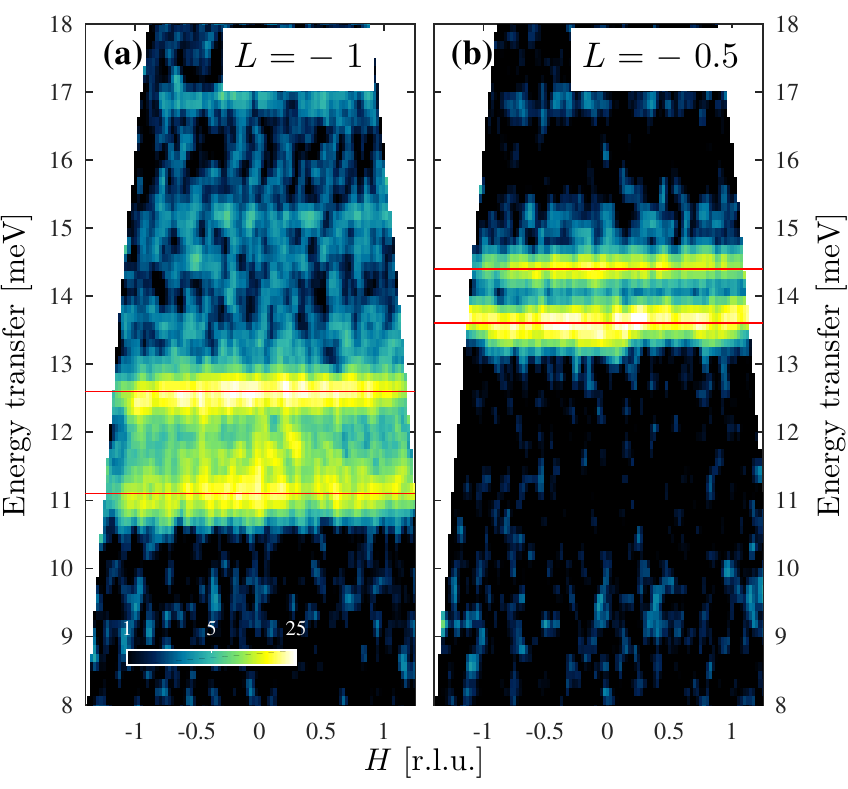}
\caption{\label{figs1} Dynamical structure factors, $S(\vec{Q},\omega)$, 
measured at 4 K for scattering vectors perpendicular to the chain direction. 
The intensity is integrated over the windows (a) $- 1.05 < L < - 0.95$ 
r.l.u.~and (b) $- 0.55 < L < - 0.45$ r.l.u. The red lines are guides to the 
eye.}
\end{figure}

To justify the statement made in the main text that RbCoCl$_3$ is a 
very one-dimensional (1D) magnetic system, in Fig.~\ref{figs1} we show 
representative cuts of our scattered intensity data for $\vec{Q}$ in the 
$H$ direction with two choices of $L$. The broad lower and narrow upper 
excitations correspond respectively to the continuum and lowest bound-state 
features visible in Fig.~4 of the main text. The extremely flat dispersion in 
both normal directions is best modelled by a constant energy, i.e.~any $H$- or 
$K$-dependence of the dispersion is smaller than the instrumental resolution.

\subsection{S2. Cluster Heat Bath Monte Carlo Simulations}

\begin{figure}[b]
\includegraphics[width=8cm]{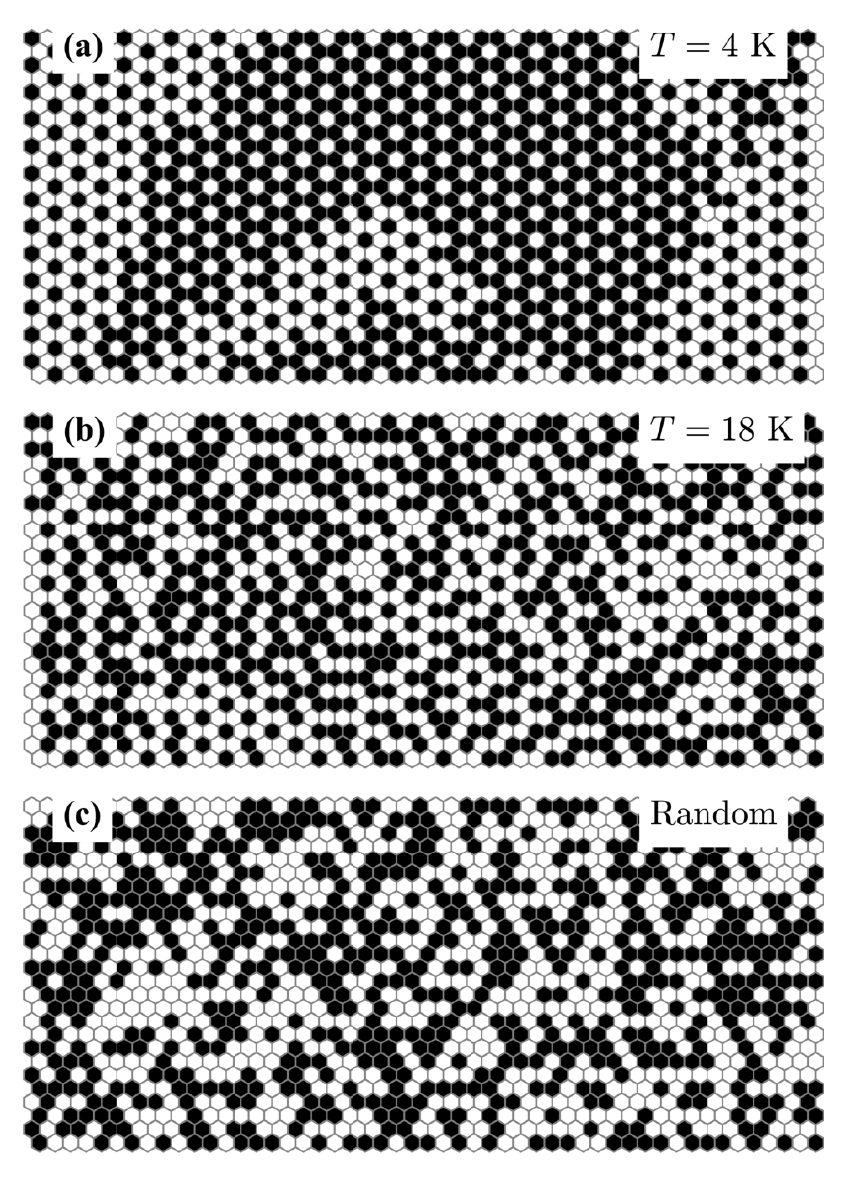}
\caption{\label{figs2} Snapshots of spin configurations obtained from CHB 
simulations at temperatures corresponding to (a) 4 K and (b) 18 K in RbCoCl$_3$.
Each hexagon represents a site in a single plane of the triangular lattice of 
antiferromagnetically coupled Ising chains, which have in addition a weak 
ferromagnetic next-nearest-neighbor interaction. Black and white colors 
represent opposing directions of the staggered Ising order. (c) Fully random 
chain configuration.}
\end{figure}

The Cluster Heat Bath (CHB) algorithm is a Monte-Carlo method that was 
developed specifically for simulating quasi-1D Ising compounds. In contrast 
to the Metropolis-Hastings algorithm, in which one spin is flipped at each 
time step, the CHB approach belongs to the broad family of cluster methods, 
where blocks of many spins may be rearranged at each step. For the coupled 
Ising system, entire spin chains are flipped to achieve a configuration where 
every chain is in equilibrium (according to the Boltzmann distribution) with 
its environment at each step. This method has been used to reproduce closely 
the magnetic phase transitions of CsCoBr$_3$ and CsCoCl$_3$ and in the present 
study we have followed the work of Refs.~\cite{matsubara1997,koseki1997,
koseki2000}. 

Here we use CHB simulations in order to gain qualitative insight into and 
semi-quantitative comparisons with our experimental results. Specifically, 
we wish to illustrate the qualitative nature of the two ordered phases, 
shown in the insets of Figs.~2(c) and 3(c) of the main text, and to verify 
quantitatively the population factors, obtained by fitting the dynamical 
structure factor at each temperature, of chains subject to the different 
possible effective staggered fields. The CHB spin structure can also be 
used to simulate a quasi-elastic scattering signal for comparison with 
experimental observations \cite{haenni2017}, which show both sharp and 
broad temperature-dependent components. We have performed simulations, using 
the interaction parameters of the main text augmented by the value $J_{nnn} = 
- J_{nn}/10$, on lattices of 120$\times$120$\times$5000 spins for up to 8000 
steps. During even-numbered steps, 10 conventional CHB operations took place 
on randomly chosen chains (i.e.~ten individual chains were flipped), while 
on odd-numbered steps flipping operations were allowed on loops of chains 
(of different, random lengths). The temperature of the system was lowered 
during the 8000 cycles, in a manner similar to simulated annealing, until 
the target temperature was obtained. 

Figures \ref{figs2}(a) and \ref{figs2}(b) show examples of equilibrated 
spin configurations in a single plane obtained for temperatures corresponding 
respectively to 4 and 18 K in RbCoCl$_3$. These are more extended versions of 
the figures shown in the insets of Figs.~2(c) and 3(c) of the main text. It 
is clear at 4 K [Fig.~\ref{figs2}(a)] that significant domains of the system 
attain a well-ordered ``honeycomb'' pattern, in which 1/3 of the chains have a 
maximal net staggered field ($6J_{nn}$) while on 2/3 of them the field cancels. 
The contribution of other chain configurations is confined to the domain walls, 
which despite the rather weak $J_{nn}$ in RbCoCl$_3$ are relatively sparse at 
4 K. At intermediate temperatures [Fig.~\ref{figs2}(b)], the domains are 
strongly disordered and the chains are correlated laterally only over rather 
short ranges, giving a distribution in which the staggered fields $2J_{nn}$ 
and $4J_{nn}$ have significant representation, whereas $6J_{nn}$ becomes much 
less likely. For reference we show in Fig.~\ref{figs2}(c) a completely random 
planar spin configuration. 

\subsection{S3. Finite-temperature DMRG}

For a microscopic understanding of temperature effects in the Ising spin 
chain, which in the extended Matsubara framework we described by effective 
parameters $\epsilon_1(T)$ and $\Gamma(T)$, we have performed time-dependent 
DMRG calculations at finite temperatures to obtain the full spectral functions. 
The spin system is represented in a basis of matrix-product states (MPS) 
\cite{schollwoeck11,hubig17:_gener}, in which the mixed states at finite 
temperatures are represented by a purification approach where their 
grand-canonical thermal density matrices are expressed in a doubled basis 
containing auxiliary variables \cite{verstraete2004,barthel09:_spect}. We 
computed the time evolution of a spin-flip excitation using the two-site 
time-dependent variational-principle method \cite{haegeman16:_unify} in 
combination with the appropriate ``near-optimal'' auxiliary space 
transformation of Ref.~\cite{barthel13:_precis}. 

Time evolution by repeated application of matrix operators causes a growth 
in information, and one of the primary attributes of the MPS formalism is 
to allow a systematic truncation of this information to fit the available 
computational resources. Physically, the problem of entanglement growth 
sets the limits in both time and space where a correlation function may be 
calculated with acceptable numerical precision. We have performed calculations 
for chain lengths up to $L = 512$ physical sites ($1024$ including auxiliary 
sites), finding these sufficient to exclude finite-size effects in the starting 
state in all cases. At $T = 4$ K, the growth of entanglement is not significant 
on the timescale over which excitations propagate across the finite system, 
and hence it is the system size that limits the maximum attainable time.

\begin{figure}[t]
\includegraphics[width=8cm]{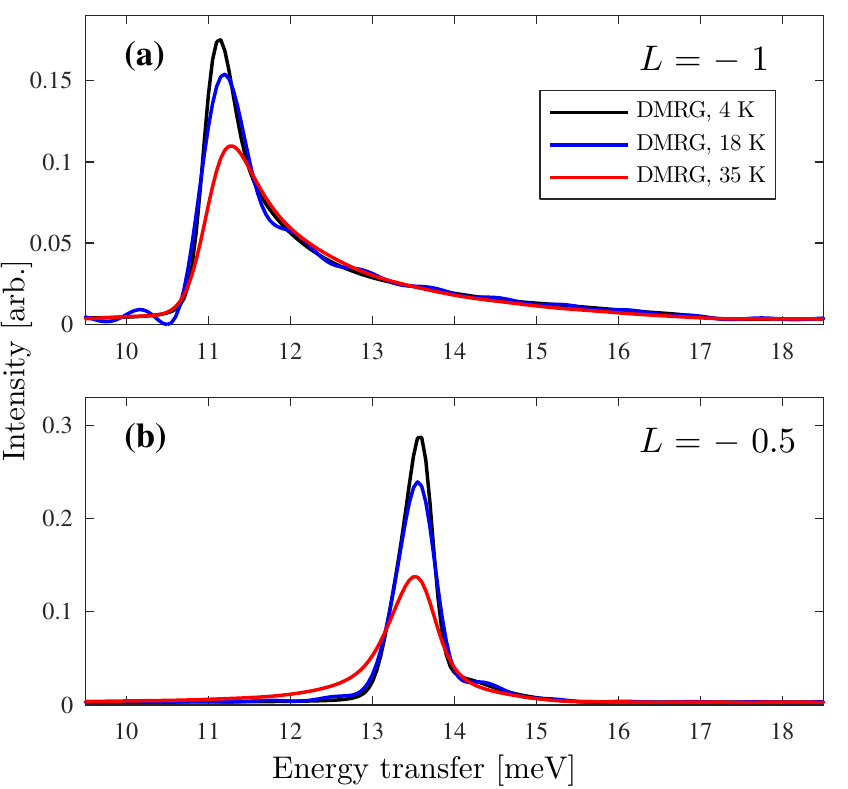}
\caption{\label{figs2p5} $S(\vec{Q},\omega)$ calculated by DMRG for the 
isolated Ising chain using the $T = 0$ interaction parameters of RbCoCl$_3$ 
with the instrumental broadening, $\sigma = 0.32$ meV, at temperatures of 4, 
18, and 35 K for (a) $L = - 1$ and (b) $L = - 0.5$.}
\end{figure}

At $T = 18$ K and $T = 35$ K, the entanglement growth is more rapid, and this 
limits the maximum attainable time to $t_{\mathrm{max}} \approx 70$ in units of 
$1/2J_1$ [Eq.~(1) of the main text], where we set the criterion of numerical 
precision to be a maximal discarded weight of $\chi^2 = 10^{-8}$. However, one 
may then extend the computed data in time, for which the linear prediction 
method of Ref.~\cite{barthel09:_spect} is particularly well suited when the 
finite temperature induces an exponential decay of the real-time correlators, 
as in the present problem. Optimizing the parameters of this interpolation 
scheme allowed us to stabilize the linear prediction for all momentum values 
and hence to reach large effective times, $t_{\mathrm{max}}^{\mathrm{pred}} \approx 
2000$. We then evaluated the dynamical response function, $S(\vec{Q},\omega,
T)$, by two Fourier transformations of the DMRG correlation functions in real 
space and time.

Figure \ref{figs2p5} illustrates the results of our DMRG calculations for 
the case of an isolated chain (zero staggered field) at temperatures of 
4, 18, and 35 K. The parameters $J_1$, $J_2$, and $\epsilon_2$ are as given 
in the main text and Table S1, and we have included a broadening equivalent 
to the instrumental resolution of 0.32 meV. We stress that $\epsilon_1 = 0.126$ 
meV is fixed to its low-temperature value in all cases, i.e.~it is not being 
changed as it is in our Matsubara procedure. Thus the fact that the peak in 
the continuum contribution at $L = - 1$ moves up in energy with increasing 
temperature [Fig.~\ref{figs2p5}(a)], signalling an effective band-narrowing, 
is an intrinsic consequence of the DMRG calculations capturing the effects 
of increased scattering processes involving thermally excited domain walls 
that are fully dynamical. We note that this intrinsic band-narrowing is 
smaller than the effective one providing optimal fits in the Matsubara 
framework, pointing to the need for a deeper investigation of dynamical 
domain-wall processes that lies beyond the scope of our present study. The 
$L = - 0.5$ peak [Fig.~\ref{figs2p5}(b)] is located near the band center 
and shows only a very weak downward trend with increasing temperature. 

On the technical side, the weak oscillation in the 18 K data is a numerical 
artifact of the limited $t_{\mathrm{max}}$ that is often suppressed by introducing 
a large effective broadening (analogous to the parameter $\Gamma$ in the main 
text). Here we have investigated different broadening schemes, but guided by 
the physics we maintain the instrumental broadening, $\sigma = 0.32$ meV. In 
this way we do not obscure the intrinsic dynamical response, particularly the 
discrete Zeeman-ladder peaks in finite staggered fields shown in Figs.~2 and 3 
of the main text and repeated in Sec.~S4. As a consequence, the oscillation in 
the 18 K data in Fig.~\ref{figs2p5}(a) is not eliminated completely, and its 
presence contributes percent-level errors in our DMRG intensity fits (Table S2) 
at this temperature; however, the line shape of the 35 K data takes its 
intrinsic form and this is important to our estimate of interchain 
correlations above $T_{N1}$ (Sec.~S5). 

\begin{figure}[t]
\includegraphics[width=8.78cm]{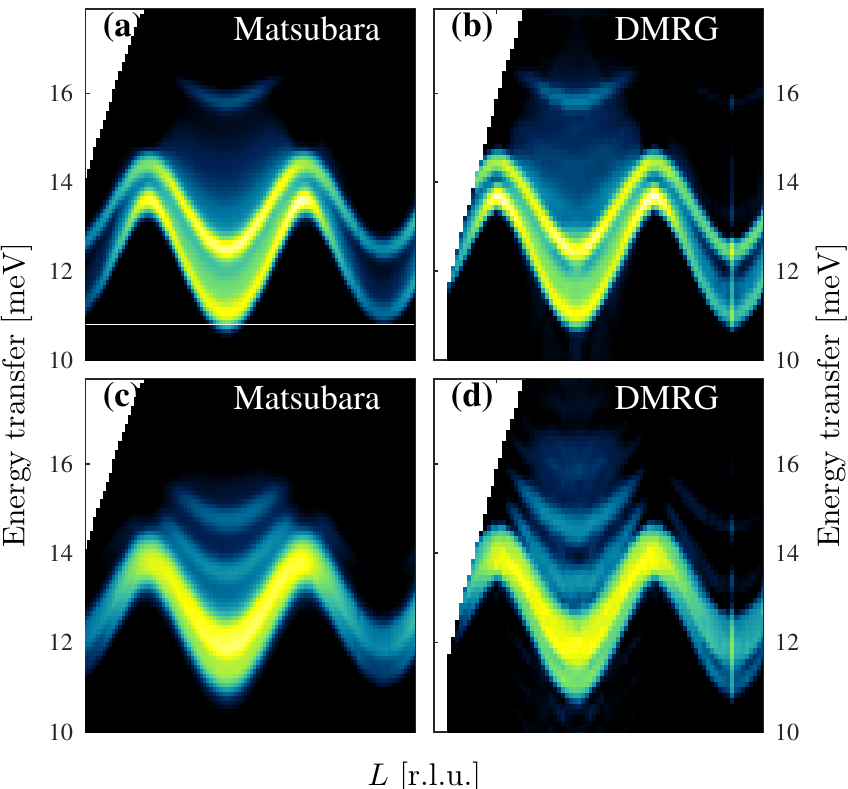}
\caption{\label{figs3} Comparison between calculations of the dynamical 
structure factor, $S(\vec{Q},\omega)$, performed within the extended 
Matsubara formalism (a,c) and by DMRG (b,d). Results are shown for the 
two temperatures at which detailed experimental data were gathered in 
the two ordered phases, namely 4 K (a,b) and 18 K (c,d).} 
\end{figure}

\subsection{S4. Data and Fitting Comparisons: Ordered Phases}

Here we show in full detail the comparison between our calculations within 
the extended Matsubara formalism and by DMRG, as well as the comparison of 
both with the spectral data measured at low ($T < T_{N2}$) and intermediate 
temperatures ($T_{N2} < T < T_{N1}$). The unique feature of the ACoX$_3$ 
materials is that, in both the fully ordered and the partially disordered 
antiferromagnetic phases, they allow the investigation of Ising chains 
subject to different effective staggered fields within a single material. 
This treatment assumes that the individual Ising chains remain as 
well-ordered clusters up to $T_{N1}$, leading to coherent effective 
staggered fields even as interchain correlations are weakened by the 
rising temperature. In Sec.~S5 we will demonstrate that this assumption, 
which is also exploited in CHB simulations, is well justified at 18 K. By 
contrast, at $T > T_{N1}$ the loss of chain order due to thermal domain-wall 
formation becomes significant and an alternative treatment is required. 

In Fig.~\ref{figs3} we show the full spectra obtained by extended Matsubara 
and by DMRG calculations. We note that Figs.~\ref{figs3}(a) and \ref{figs3}(d) 
are the same, respectively, as Figs.~2(b) and 3(b) of the main text. While it 
is not surprising that both calculations reproduce the 4 K data rather well 
[Figs.~\ref{figs3}(a) and \ref{figs3}(b)], this does demonstrate both that 
the Matsubara framework captures the primary physics of the system and that 
the DMRG methods are well within their numerical capabilities. 

\begin{figure}[t]
\includegraphics[width=8.4cm]{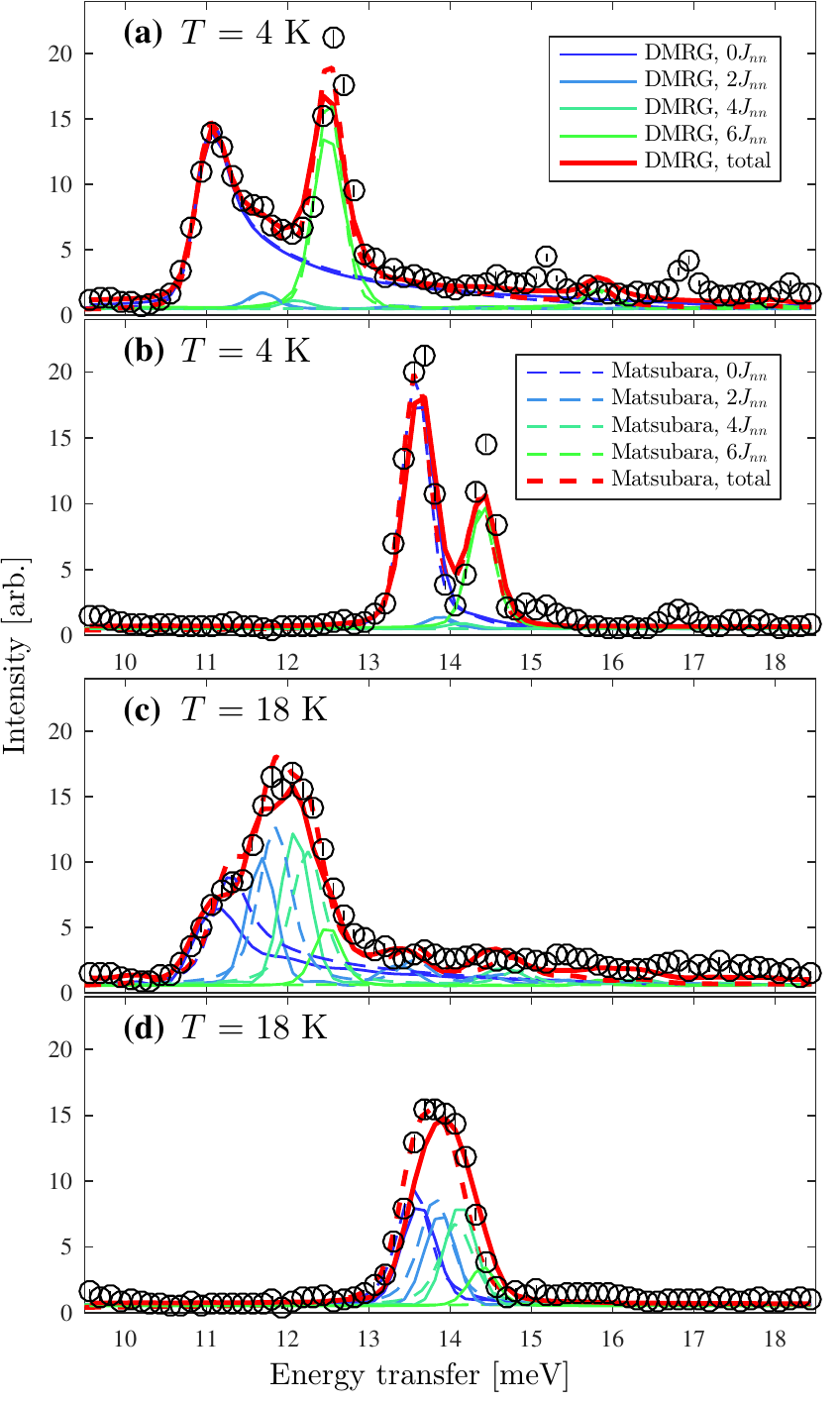}
\caption{\label{figs4} Staggered-field contributions to the dynamical 
structure factor, $S(\vec{Q},\omega)$. Measured intensities (points) and 
those calculated using both the extended Matsubara model (dashed lines) and 
DMRG (solid lines) are integrated over the $\vec{Q}$ windows $- 1.05 < L <
 - 0.95$ r.l.u.~(a,c) and $- 0.55 < L < - 0.45$ r.l.u.~(b,d). Results are 
shown at temperatures of 4 K (a,b) and 18 K (c,d). The lines changing from 
blue to green indicate the individual contributions of chains subject to 
each of the possible staggered fields, $m J_{nn}$ with $m = 0$, 2, 4, or 6, 
and the red lines show their sum.} 
\end{figure}

At 18 K, in the partially disordered antiferromagnetic phase, the level of 
agreement is non-trivial. In the extended Matsubara formalism, the fit to the 
data can, as discussed in the main text, be achieved by allowing only two of 
the six parameters to have a temperature-dependence, one affecting the band 
width and one the line width. By contrast, in DMRG there are no free parameters 
and both effects are intrinsic. Table \ref{tabs1} shows a comparison between 
the fitting results we obtain from the two procedures: there is close agreement 
on $J_1$ and $J_{nn}$, some discrepancy in the next-neighbor parameters $J_2$ 
and $\epsilon_2$, and of course the difference in treatment of the effective 
band width, contained in $\epsilon_1$. 

\begin{table}
\begin{center}
\begin{tabular}{ c || c | c }
  & \;\; Matsubara \;\; & \;\; DMRG \;\; \\
\hline
$J_1$ [meV] & 5.89 & 5.86 \\
$J_2$ [meV] & $-0.518$ & $-0.576$ \\
$J_{nn}$ [meV] & 0.129 & 0.128 \\
\hline
$\epsilon_1$ at 4 K & 0.126 & 0.126 \\
\;\; $\epsilon_1$ at 18 K \;\; & 0.112 & 0.126 \\
$\epsilon_1$ at 35 K & 0.101 & 0.126 \\
\hline
$\epsilon_2$ & 0.605 & 0.559 \\
\end{tabular}
\caption{\label{tabs1} Comparison of fitting parameters obtained from the 
extended Matsubara Hamiltonian and from DMRG calculations at 4 K and 18 K. 
Error bars are omitted.}
\end{center}
\end{table}

\begin{table}
\begin{center}
\begin{tabular}{ c | c || c | c | c | c }
Method & \;\; $T$ [K] \;\;  & \;\; $I_{0}$ \;\; & \;\; $I_{2}$ \;\; & \;\; 
$I_{4}$ \;\; & \;\; $I_{6}$ \;\; \\
\hline
\;\; Matsubara \;\; & \multirow{3}{*}{4} & 67(3) & 3(3) & 0(3) & 31(3) \\
DMRG & & 65(3) & 3(3) & 1(3) & 31(3) \\
CHB & & 61(1) & 6(1) & 5(1) & 28(1) \\
\hline
Matsubara &\multirow{3}{*}{18} & 39(5) & 35(5) & 27(5) & 0(5) \\
DMRG & & 35(5) & 27(5) & 28(5) & 10(5) \\
CHB & & 39(1) & 35(1) & 20(1) & 6(1) 
\end{tabular}
\caption{\label{tab2} Comparison of chain population factors deduced from 
extended Matsubara and DMRG calculations at 4 K and 18 K, and from CHB 
simulations on a system of 120$\times$120 chains. $I_m$ is the percentage 
of the scattered weight that may be ascribed to a chain in a staggered 
field $h = m J_{nn}$. The sum of the intensities may deviate from 100\% due 
to rounding effects. Because the measured $S(\vec{Q},\omega)$ datasets contain 
many thousands of points, statistical errors in the fitting procedure are of 
order 0.1\%. The quoted error bars for Matsubara and DMRG results represent 
the estimated systematic uncertainties in the fitting process, and for CHB 
in the simulations.} 
\end{center}
\end{table}

These results are predicated on two subsidiary calculations, namely the 
intensity contributions due to chains in the different effective staggered 
fields and the weights of each chain type in the final sum; we take only the 
latter as free parameters. To show clearly the contributions of the different 
types of chain (i.e.~different staggered fields) to the measured intensity, we 
have performed our calculations separately for staggered fields of 0, $2J_{nn}$, 
$4J_{nn}$, and $6J_{nn}$. The results, displayed in Figs.~\ref{figs4}(a,b) and 
\ref{figs4}(c,d), show in full detail the respective panels of Figs.~2(c,d) 
and 3(c,d) of the main text. Both the energy levels of the Zeeman ladders 
and the corresponding intensities computed within the extended Matsubara 
description and by DMRG are equal with quantitative accuracy at 4 K. 
However, it is clear at 18 K that the DMRG results do not contain as much 
narrowing of the band width as that optimizing the Matsubara fits, as a 
result of which the Zeeman-ladder states are less strongly renormalized and 
the weight factors appropriate to reproduce the measured intensity, shown 
in Table \ref{tab2}, include stronger contributions from higher $m$. 
Nonetheless, these discrepancies lie close to the systematic error bars 
on the fitted $I_m$ percentages, and both fits are consistent with our CHB 
simulations (Sec.~S2). One may certainly conclude that the treatment of 
neighboring chains in terms of an effective staggered field does provide 
an accurate reflection of the response of the 3D system in its ordered phases.

\subsection{S5. Data and modelling above $T_{N1}$}

For the discussion in Sec.~S4 we excluded the high-temperature regime. In the 
magnetically disordered state, it is not appropriate to use a formalism based 
on effective staggered magnetic fields, i.e.~on long segments of uniform 
interchain order. Nor, however, is it appropriate to neglect all interchain 
interactions at temperatures close to but above $T_{N1}$ (Fig.~4(a) of the main 
text). In this regime, thermally induced randomness appears both in the chains, 
in the form of single thermal domain walls, and between the chains, in a form 
that can be modelled by an increasingly random effective field. In the 
Matsubara framework, a single domain wall acts to terminate each chain 
segment and we model the first effect by considering a static thermal 
distribution of domain walls, and hence of chain segment lengths, within 
a Monte Carlo approach. By applying the same approach to the neighboring 
chains, we also account for the second effect. Here it is important to note 
that our DMRG calculations, which are performed for a single chain at finite 
temperatures, are computing the first of these two effects for fully dynamic 
thermal domain walls. 

\begin{figure}[t]
\includegraphics[width=8cm]{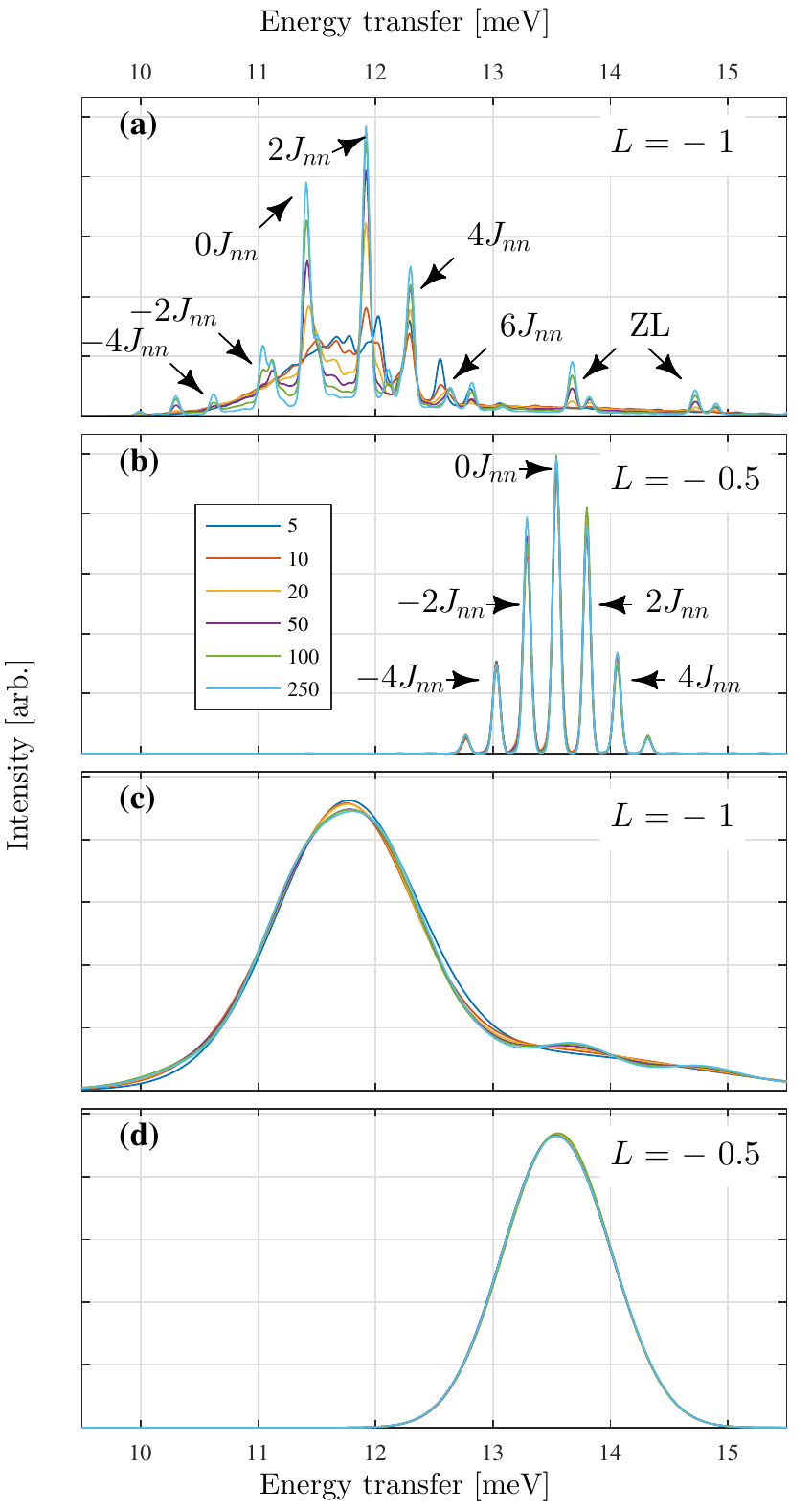}
\caption{\label{figs6} Illustration of scattered intensities obtained by 
Monte Carlo modelling within the Matsubara framework for coupled chains in a 
thermally random effective field with the average length, $\langle n \rangle$, 
of the Ising chain segments as the adjustable parameter. All parameters are 
those of the Matsubara fit with effective band-width parameter $\epsilon_1 = 
0.101$. (a) $L = - 1$, $\sigma = 0.03$. (b) $L = - 0.5$, $\sigma = 0.03$. (c) 
$L = - 1$, $\sigma = 0.32$. (d) $L = - 0.5$, $\sigma = 0.32$. The lowest modes 
of each Zeeman ladder are marked in panels (a) and (b) by ``$mJ_{nn}$'' and ZL 
denotes higher Zeeman-ladder modes in panel (a).}
\end{figure}

In Figs.~\ref{figs6}(a) and \ref{figs6}(b) we illustrate the results of the 
Matsubara-based procedure for the parameters of RbCoCl$_3$ with a very low 
broadening ($\sigma = 0.03$ meV). For long average segment lengths, $\langle 
n \rangle$, at $L = - 1$ [Fig.~\ref{figs6}(a)] the isolated-chain continuum 
and the Zeeman-ladder peaks for all three finite values of $m$ are clearly 
discernible, along with faint signals for $-m$. At $L = - 0.5$ 
[Fig.~\ref{figs6}(b)], where the modes are non-dispersive, the intensities 
of the lowest modes of each Zeeman ladder, which have separation $2J_{nn}$, 
approach a 1:6:15:20:15:6:1 distribution, while the higher modes of all 
ladders are very weak. This regime is the basis on which, by inspection of 
the data at 18 K (Fig.~3 of the main text), one may conclude that the approach 
of effective staggered fields remains well justified at that temperature. As 
the density of thermal domain walls increases, i.e.~as $\langle n \rangle$ 
decreases, it is clear at $L = - 1$ that the low-$T$ features lose weight 
and that scattered-intensity contributions appear at many different energies 
as chain segments of all possible lengths contribute (including those of only 
one and two spins). However, at $L = - 0.5$ all of these segments continue to 
contribute at the same energies. 

\begin{figure}[t]
\includegraphics[width=8.2cm]{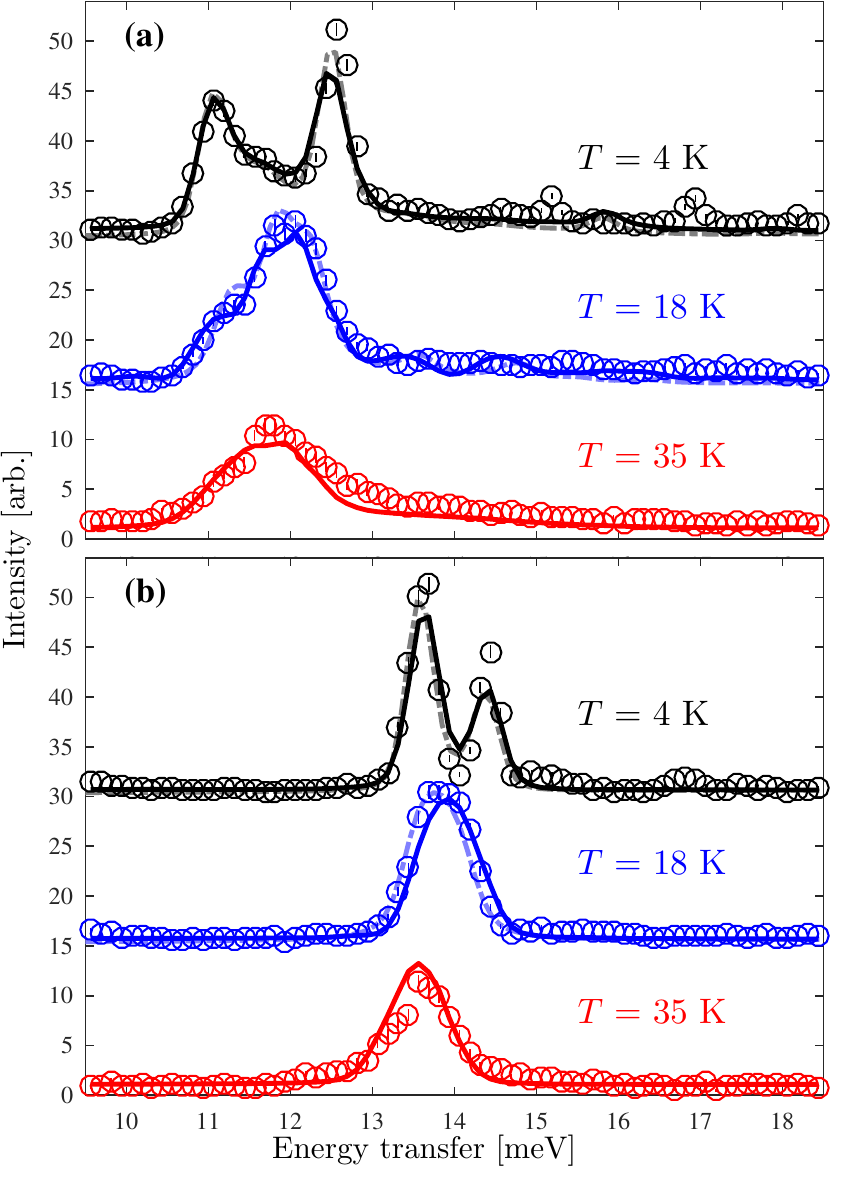}
\caption{\label{figs5} Thermal evolution of $S(\vec{Q},\omega,T)$, shown by 
superposing datasets taken at all three temperatures. Measured intensities 
(points) and those calculated using both the extended Matsubara model (dashed 
lines) and DMRG (solid lines) are integrated over the $\vec{Q}$ windows 
(a) $- 1.05 < L < - 0.95$ r.l.u.~and (b) $- 0.55 < L < - 0.45$ r.l.u., and 
shown with an offset of 15 for clarity. Solid lines show the fits discussed 
in Secs.~S4 and S5.}
\end{figure}

To model our 35 K data, we first restore the instrumental resolution, 
$\sigma = 0.32$ meV. As Figs.~\ref{figs6}(c) and \ref{figs6}(d) make clear, 
this causes all of the separate features of the response at any average 
segment length to merge into a single, broad peak. Somewhat surprisingly, the 
shape of this feature, which at $L = - 1$ is centered between the energies of 
the $m = 0$ and $m = 2$ peaks, becomes independent of $\langle n \rangle$. 
Thus although we cannot relate $\langle n \rangle$ directly to the temperature 
of the system, its effects become irrelevant due to the combined effects of 
the ``splitting'' $2J_{nn}$ and the instrumental broadening, which establish 
the width of the broad feature. Its position is controlled by the effective 
band-width parameter, $\epsilon_1$, which may therefore be fixed with 
reasonable accuracy using the experimental data. Beyond the fact that the 
domain walls in our modelling procedure are static rather than dynamic, which 
is also accounted for crudely by the effective $\epsilon_1$, we expect the 
primary inaccuracy to lie in the neglected effects of $J_2$ across a thermal 
domain wall. In this sense the physics of coupled Ising-chain systems with 
thermal randomness poses a quantitative challenge to more specialized 
theoretical and numerical techniques. 

Returning again to our experiments, it was shown clearly in the main text
that, despite $T$ exceeding $T_{N1}$, the scattered intensity at 35 K is far 
from that of chains isolated from each other by strong thermal fluctuations. 
To quantify the remaining interchain correlation effects, we fit the intensity 
measured at 35 K to a weighted sum of the isolated-chain response, taken from 
DMRG at 35 K, and the response of the thermally disordered system with a 
realistic average segment length of $\langle n \rangle = 10$ sites and the 
value $\epsilon_1 = 0.101(2)$ given in Table \ref{tabs1}. In this fit we also 
include the actual energy and momentum steps of the experimental data binning, 
which is responsible for the discrepancy in shape between the smooth model of 
Fig.~\ref{figs6}(c) and the more discrete ``disordered coupled chains'' 
response in Fig.~4(a) of the main text.  
 
The results of this procedure, shown by the red lines in Figs.~4(a) and 4(b) 
of the main text, indicate that approximately 63\% of the measured response 
can be ascribed to residual interchain correlations, with a systematic error 
of order 5\%. Although this seems to be a surprisingly large fraction, it 
should be borne in mind that essentially all of the chains are correlated 
below $T_{N1}$, which is only 7 K lower, because with a random field there 
is no longer any cancellation effect of the type determining the response 
of 2/3 of the Ising chains in the FI phase. As noted in the main text, 
susceptibility measurements indicate that these correlations persist up to 
temperatures around 80 K in RbCoCl$_3$ while diffuse scattering measurements 
\cite{haenni2017} confirm their presence up to 60 K.

We now step back to consider the physics of the system. In Fig.~\ref{figs5} 
we illustrate the evolution of the dynamical structure factor with temperature 
by comparing the zone-center and zone-edge intensities at all three measurement 
temperatures. This highlights the rapid loss of the bound-state 
(staggered-field) contributions, the broadening of both the continuum 
and the remaining bound-state signals, and the upshift of the lower part of 
the band that can be understood as a narrowing effect due to the scattering 
of propagating domain-wall pairs on thermally excited domain walls. 
Figure \ref{figs5} allows a clear visualization of the way in which the 
Matsubara (domain-wall) description allows these changes to be captured by 
only two thermal parameters (for line width and band width) and highlights 
the power of state-of-the-art DMRG methods to compute the full response of 
1D systems at finite temperatures and energies. 

For perspective on our modelling of the 3D Ising system, the Matsubara 
and DMRG methods are complementary in that DMRG provides the fundamental 
strongly correlated quantum physics, albeit at significant computational 
expense, which makes it difficult to perform iterative fits of experimental 
data and prohibitive to include randomness; the extended Matsubara (effective 
Hamiltonian) framework is cheap and easy to apply for iterative fitting, but 
its approximate inclusion of thermal effects requires a benchmark that DMRG 
can provide.

\end{document}